\documentclass[]{ar2e-changed}

\usepackage{amsmath,amssymb}
\usepackage{latexsym}
\usepackage{graphicx}
\usepackage{dcolumn}
\usepackage{bm}
\usepackage{psfrag}
\usepackage{subfigure}
\usepackage{tabularx}

\def\LSCO{La$_{2-x}$Sr$_x$CuO$_4$}

\def\YBCO{YBa$_2$Cu$_3$O$_{6+x}$}

\def\C60{A$_x$C$_{60}$}
\def\LNSCO{La$_{1.6-x}$Nd$_{0.4}$Sr$_x$CuO$_{4}$}

\def\SROone{ Sr$_{2}$Ru$_{}$O$_{4}$}
\def\SROtwo{ Sr$_{3}$Ru$_{2}$O$_{7}$}

\def\BSCCO{Bi$_2$Sr$_2$CaCu$_2$O$_{8+\delta}$}
\def\oxychloride{Ca$_{2-x}$Na$_x$CuO$_2$Cl$_2$}
\def\LNSCO{La$_{1.6-x}$Nd$_{0.4}$Sr$_x$CuO$_{4}$}

\def\HgCu3{HgCa$_2$Cu$_3$O$_{8+y}$}
\def\HgCu4{HgBa$_2$Ca$_3$Cu$_4$O$_{10+y}$}
\def\TlCu{Tl$_2$Ba$_2$CuO$_{6+\delta}$}
\def\TlCu3{Tl$_2$Ba$_2$Ca$_2$Cu$_3$O$_{10+y}$}
\def\TlCu4{Tl$_2$Ba$_2$Ca$_3$Cu$_4$O$_{12+y}$}

\def\BiCu3{Bi$_2$Sr$_2$Ca$_{2}$Cu$_3$O$_y$}
\def\BiCaMnO{Bi$_{1-x}$Ca$_x$MnO$_3$}

\def\8LSCO{La$_{1.88}$Sr$_{.12}$CuO$_4$}
\def\110LNSCO{La$_{1.5}$Nd$_{0.4}$Sr$_{0.1}$CuO$_{4}$}
\def\stage4LCO{La$_{2}$CuO$_{4+\delta}$}
\def\Y248{YBa$_2$Cu$_4$O$_8$}

\def\NbSe2{NbSe$_2$}
\def\TaSe2{TaSe$_2$}
\def\TiSe2{TiSe$_2$}
\def\NaCoOH2O{Na$_{0.3}$CoO$_{2y}$H$_2$O}
\def\MgB2{MgB${}_2$}

\def\URu2Si2{URu$_2$Si$_2$}

\def\hts{high temperature superconductors}

\newcommand{\xdir}{$\langle 1\overline{1} 0 \rangle$}
\newcommand{\ydir}{$\langle 110 \rangle$}
\newcommand{\zdir}{$\langle 001 \rangle$}

\begin{document}

\title{Nematic Fermi Fluids in Condensed Matter Physics}
\markboth{Fradkin, Kivelson, Lawler, Eisenstein \& Mackenzie}{Nematic Fermi Fluids in Condensed Matter Physics}

\author{Eduardo Fradkin
\affiliation{Department of Physics, University of Illinois, Urbana, Illinois 61801-3080; efradkin@illinois.edu}
Steven A. Kivelson
\affiliation{Department of Physics, Stanford University, Stanford, California 94305-4060}
Michael J. Lawler
\affiliation{Department of Physics, Applied Physics \& Astronomy, Binghamton University, Binghamton, New York 13902, USA, and 
Department of Physics, Cornell University, Ithaca, New York 14853, USA}
James P. Eisenstein
\affiliation{Condensed Matter Physics, California Institute of Technology, Pasadena, California 91125}
Andrew P. Mackenzie
\affiliation{Scottish Universities Physics Alliance, School of Physics and Astronomy,  University of St. Andrews, North Haugh, St. Andrews, Fife KY16  9SS, United Kingdom}
}


\begin{abstract}
Correlated electron fluids 
 can exhibit a startling array of complex phases, among which one of the more surprising is the electron nematic, a translationally invariant metallic phase with a spontaneously generated spatial anisotropy.  Classical nematics generally occur in liquids of rod-like molecules;  given that electrons are point like, the initial theoretical motivation for contemplating electron nematics came from thinking of the electron fluid as a quantum melted electron crystal, rather than a strongly interacting descendent of a Fermi gas. Dramatic transport experiments in ultra-clean quantum Hall systems in 1999 and 
in Sr$_3$Ru$_2$O$_7$ in a strong magnetic field in 2007 established that such phases exist in nature.
 In this article, we  
 briefly review the theoretical considerations governing nematic order, summarize the quantum Hall and Sr$_3$Ru$_2$O$_7$ experiments that unambiguously establish the existence of this phase, and survey some of the current evidence for such a phase in the cuprate and Fe-based {\hts}.  
\end{abstract}
\date{\today}
\maketitle

\section{Introduction}
\setcounter{footnote}{0}

Strongly correlated electron systems are mostly defined by what they are not:  they are not gases of weakly interacting quasiparticles (QPs) (Fermi gases), but they are still electron fluids  ({\it i.e.} not insulators).  
Perhaps classical liquids are a good analogy. They are so different locally from a gas that the two phases are usually separated by a strongly first order phase transition. 
In addition, the phase diagrams of complex classical fluids contain a rich set of liquid crystalline phases that exhibit varying degrees of translation and rotational symmetry breaking. Alas, it is nearly as difficult to obtain a satisfactory understanding of classical liquids as of highly correlated quantum fluids.  Nevertheless, many spectacular thermodynamic and hydrodynamic properties follow directly from the symmetry breaking in liquid crystalline phases \cite{degennes-1993,chaikin-1995}.

Research in strongly correlated electronic systems during the past decade has revealed the existence of a variety of novel phases of quantum matter,
electronic liquid crystal phases. In particular the simplest of 
these phases, the nematic Fermi fluid, has received a great deal of attention. Transport measurements in two-dimensional electron systems (2DES) at high magnetic fields \cite{lilly-1999,du-1999}, strontium ruthenate materials \cite{borzi-2007} and in several {\hts} \cite{ando-2002,hinkov-2007}, show unusually large strongly temperature-dependent transport anisotropies in these otherwise essentially isotropic electronic systems. By analogy with their classical counterparts, the homogeneous anisotropic phase of these strongly quantum mechanical systems is said to be an {\it electronic nematic phase} \cite{kivelson-1998,fradkin-1999,kivelson-2003}.

The origin of the behavior of classical complex fluids lies in the microscopic structure of the molecules ({\it i.e.\/} their rod-like shapes) and their anisotropic interactions. In their nematic phase rotational invariance is broken spontaneously and the fluid exhibits orientational order. Various sorts of smectic phases
that break translational and rotational invariance to various degrees
are also observed \cite{degennes-1993}. If the rod-like molecules are chiral ({\it i.e.\/} have a screw-like structure) more complex phases such as cholesterics and blue
phases result. Whereas liquid crystals typically have short range interactions, the related behavior of ferrofluids arises, in part, from their long-range dipolar interactions.

In contrast to their classical relatives, the quantum versions of liquid crystal phases typically arise in strongly correlated systems whose constituents are electrons and hence are point particles.
 Thus the physical origin of quantum liquid crystal phases is very different from that of classical ones. 
 Typically, various forms of {\it quantum soft matter} 
arise  as a consequence of  a form of local (often Coulomb-frustrated) electronic 
 phase separation \cite{emery-1993}. Rather than a fluid of electrons the system consists of mesoscale emulsions of locally Mott insulating regions separated by more metallic regions. 
 As a consequence, 
 these systems have rich phase diagrams whose phases exhibit a spectrum of patterns each breaking different spatial symmetries ranging from uniform fluids, typically with the standard properties of Fermi liquids \cite{baym-1991}, to complex Wigner-like solids (possibly with many electrons per unit cell), with various sorts of electronic
 liquid crystals in between, especially nematic and stripe (smectic) phases. For similar reasons, pasta phases (analogues of the smectic phase)
 were proposed 
 to arise in the crusts of neutron stars (lightly doped with protons)  
 to explain the glitches observed in gamma-ray emissions \cite{ravenhall-1983,ravenhall-1993}.

One important and unique feature of electronic liquid crystals \cite{kivelson-1998}
is that  other types of 
order, including magnetic and superconducting order, 
can be {\it intertwined} with the liquid crystalline order. 
 This leads to an array of new phases
 such as stripe phases
  \cite{tranquada-1995,kivelson-2003,vojta-2009}, and  the striped superconductor \cite{berg-2008a,berg-2009,berg-2009b}
  in which the charge, spin and superconducting orders all have smectic character.
Moreover, a host of 
 phases are possible in which rotational invariance in real space and in spin space are broken simultaneously. An example of such a state is the nematic-spin-nematic phase,\cite{kivelson-2003} and its generalizations\cite{wu-2004,wu-2007}.  A rich set of phases of this type is discussed in Reference \cite{wu-2007}. There are interesting conceptual connections between these more general liquid crystal phases and hidden orders that have been suggested in the context of the cuprate superconductors\cite{chakravarty-2001c,varma-2005} and heavy-fermion systems \cite{varma-2006}.

In this review we will discuss the physics of the electronic nematic phase. In the next section, we define and characterize the broken symmetries, and 
discuss how point-like electrons can nevertheless exhibit nematic order. The following three sections then focus on dramatic realizations of these phases in quantum Hall systems, Sr$_3$Ru$_2$O$_7$ and {\hts}. We conclude with an outlook on future directions with an emphasis on new systems that may exhibit nematic order and open questions.

\section{General Theoretical Considerations}
\setcounter{footnote}{0}
\subsection{Definitions}

We use the term electron nematic to denote an electron fluid that spontaneously breaks a symmetry of the underlying 
Hamiltonian which interchanges two axes of the system. 
In its pure form, a nematic phase does not spontaneously break inversion symmetry, time reversal symmetry, or translational 
symmetry. However, we sometimes use the term when one of those additional symmetries is  explicitly broken. For instance, we 
talk about nematic phases in the presence of an applied 
 magnetic field that explicitly breaks time reversal symmetry.
That the nematic phase  is a fluid implies
 that it is  conducting (metallic) or superconducting. 

In the simple case of a crystal with a four-fold rotational symmetry,
 the nematic phase breaks the $C_4$ symmetry to a residual $C_2$. 
 A nematic phase can also occur when reflection symmetry through a plane is spontaneously broken.  Since the spatial 
 symmetries of a crystal are discrete, this nematic phase always leads to a two-fold degenerate ground-state, and hence it is 
 sometimes referred to as an Ising nematic. In some cases, however, crystal field effects are weak and the system may have 
 an approximate continuous rotational symmetry.  In two dimensions or quasi-2D systems, the nematic state then breaks a 
 continuous rotational symmetry, $C_\infty$, to $C_2$;  this we refer to as an XY-nematic phase.
For instance, since the full rotational symmetry of electrons near the $\Gamma$ point in a semiconductor is only broken by higher 
order corrections to the effective mass approximation, nematic phases in such circumstances (including the quantum Hall 
nematic) are likely to be approximately XY nematics.
For systems with full rotational symmetry, it is reasonable to consider more complex patterns of rotation symmetry 
breaking, such as an electron hexatic phase which would break $C_\infty$ to $C_6$.

As is so often the case, there are many ways of looking at a given pattern of broken symmetry, with different names associated with different presumptions concerning the underlying physics.  In this case, transitions in which the point group symmetry of a crystal alters have been known for decades.  A metal that undergoes a transition from a tetragonal to an orthorhombic or from an orthorhombic to a monoclinic crystal structure has undergone a nematic transition according to the above 
definition.  In many cases, when the transition is driven largely by structural forces having little to do with the low energy electronic degrees of freedom, there is no insight to be drawn from the electronic liquid crystalline perspective.  However, when the driving force for the symmetry change comes from interesting electronic physics, and especially where the effects of the symmetry breaking are much more pronounced on the electronic structure than on the crystalline structure, the liquid crystalline perspective is appropriate.  Nevertheless, 
symmetry breaking in the electron fluid necessarily implies symmetry breaking in the crystal structure and vice versa.  

\subsection{Symmetry Considerations}

Many features of the physics of the nematic phase are dictated by symmetry.  
In most cases in which electron nematic phases arise, the Landau theory admits only even-order invariants, and thus whether the transition is continuous or discontinuous involves microscopic considerations. Clearly, the transition to an Ising nematic phase, if continuous, is in the Ising universality class whereas for an XY-nematic, it is XY-like, e.g. the transition is Kosterliz-Thouless-like in two dimensions.  
In the ordered phase, an XY-nematic will possess a Goldstone mode associated with the broken symmetry.

The order parameter of a nematic phase is properly represented as a traceless symmetric tensor, ${\cal N}_{ab}$. Thus
any measurable traceless symmetric tensor physical property can represent ${\cal N}_{ab}$.  As an example, consider the case in which the nematic order is associated with the breaking of $C_4$ rotational symmetry in the x-y plane.  (For a 2D system, this is just the plane of the system, whereas for a 3D non-cubic crystal, this is an appropriately chosen symmetry plane.)  In this case, choosing x and y to be the principle axes of the nematic phase, possible order parameters are:
\begin{equation}
\mathcal{N}=\frac { \rho_{xx} - \rho_{yy}} { \rho_{xx}+\rho_{yy} }
\ \ \ {\rm or} \ \ \
\mathcal{N}=\frac { S( \vec Q) - S(\vec Q^\prime)} { S(\vec Q) + S(\vec Q^\prime)} 
\ \ \ {\rm or} \ \ \
\mathcal{N}= \frac {\langle \vec \sigma_{\vec R}\cdot \vec \sigma_{\vec R+\hat x}\rangle - \langle \vec \sigma_{\vec R}\cdot \vec \sigma_{\vec R+\hat y} \rangle}
{\langle \vec \sigma_{\vec R}\cdot \vec \sigma_{\vec R+\hat x}\rangle + \langle \vec \sigma_{\vec R}\cdot \vec \sigma_{\vec R+\hat y} \rangle}
\label{N}
\end{equation}
where $\rho_{ab}$ is the resistivity tensor, $S(\vec Q)$ is a convenient structure factor with $\vec Q= Q \hat x$ and $\vec Q^\prime=Q\hat y$ being any symmetry-related vectors along the symmetry axes, and $\vec\sigma_{\vec R}$ is a spin operator on site $\vec R$, and ${\cal N}_{ab} \equiv {\cal N}\delta_{ab}[\delta_{a,x}-\delta_{a,y}]$.  Other possible definitions of the nematic order parameter can be constructed by analogy with these.    In all cases, the order parameter is identically zero in the symmetric phase and non-zero in the nematic phase.   It is also sometimes convenient to think of the nematic order parameter as a headless vector pointing parallel or antiparallel to the preferred axis. So long as this vector lies in a fixed plane of the crystal, its direction can be specified by a single azimuthal angle, $\theta$ which is defined mod $\pi$, since it is a headless vector.

In real-world situations, it is not possible to ignore the role of quenched disorder.  Because the nematic phase involves a broken spatial symmetry, the degeneracy between the two possible ground states is always lifted by disorder.  In the equivalent Ising or XY model, this means that disorder appears as a random field, that is,  it produces a term in the effective Hamiltonian of the form
\begin{equation}
H_{dis}=\int d\vec r \  {\rm Tr}[ \ h(\vec r){\cal N}(\vec r)\ ]
\label{RFIM}
\end{equation}
where $h$ is a quenched, random tensor quantity with zero mean which reflects the local configuration of the disorder.
Note that the trace of $h$ (which represents the scalar disorder) does not couple to the nematic order.  In more physical terms the value of the disorder potential at a given point in space does not define a preferred direction, and thus does not serve to pin the nematic order.  However, in a patch large enough that the disorder potential varies substantially, a preferred axis can be defined in terms of appropriate spatial derivatives of the disorder potential.  
This means that smoothly varying disorder potentials, for which $h_{ab}(\vec r) \propto \partial_a\partial_b V(\vec r)$, couple relatively weakly to nematic order.  

Nonetheless, once there is any random field at all, 
 the consequences are qualitatively important, as can be seen from the precise analogy implied by Eq. \ref{RFIM} and the random-field Ising model or the random-field XY model. For an XY nematic in $D < 4$ and for an Ising nematic in $D \leq 2$, no macroscopic symmetry breaking can occur in the presence of randomness, no matter how weak the coupling to the nematic order.  Whereas for an Ising nematic in 3D, macroscopic symmetry breaking survives to a critical disorder strength, if the system is quasi 2D (as is typically the case in the materials that have shown tendencies to nematic order to date), the critical disorder strength is parametrically small.  Moreover, in the absence of symmetry breaking, one generally expects that the phase transition found in the absence of disorder will be rounded, leaving no precise criterion to distinguish the two phases.

The absence of spontaneous nematic symmetry breaking in the presence of disorder makes the identification of electron nematic phases intrinsically difficult.  One solution is to look for nematic symmetry breaking using some form of local probe:  There is a length scale, $\xi_{domain}$, which characterizes the size of nematic domains and which diverges as the strength of the random field tends to zero. Any probe that is sensitive to correlations at shorter length scales than $\xi_{domain}$ will see symmetry breaking effects that are untainted by the quenched randomness.  Moreover, in very clean samples, the local pinning of the nematic domains is very weak and, hence, a substantial degree of nematic order can be induced by the application of a relatively weak applied symmetry breaking field ({\it i.e.} by the addition of a uniform component of $h$).  Thus, an anomalously large response to a small symmetry breaking field that onsets below a well defined crossover temperature, $T^*$ (roughly equal to the transition temperature in the absence of disorder), is also a possible way to detect a nematic phase.  An attractive approach that is only beginning to be exploited
\cite{carlson-2006,weissman-2009} is to look for characteristic dynamical ({\it e.g.} noise) signatures of the presence of pinned domains;  again, where there is a nematic phase, then for weak enough disorder, glassy signatures of the presence of large nematic domains should onset fairly sharply below a well defined crossover temperature, $T^*$.

\subsection{Mechanism of electron nematic formation}

While the symmetry-related features of the nematic phase are thoroughly discussed in the standard statistical mechanics literature, microscopic features of the nematic state are much more complicated, difficult to treat theoretically, and dependent on details of the material in question.  We cover this below where we discuss specific realizations of the nematic state.  However, there is a general issue that we do address here: the mechanisms by which point-like shapeless electrons form a nematic phase. 

\begin{figure}[hbt]
\begin{center}
\includegraphics[width=\textwidth]{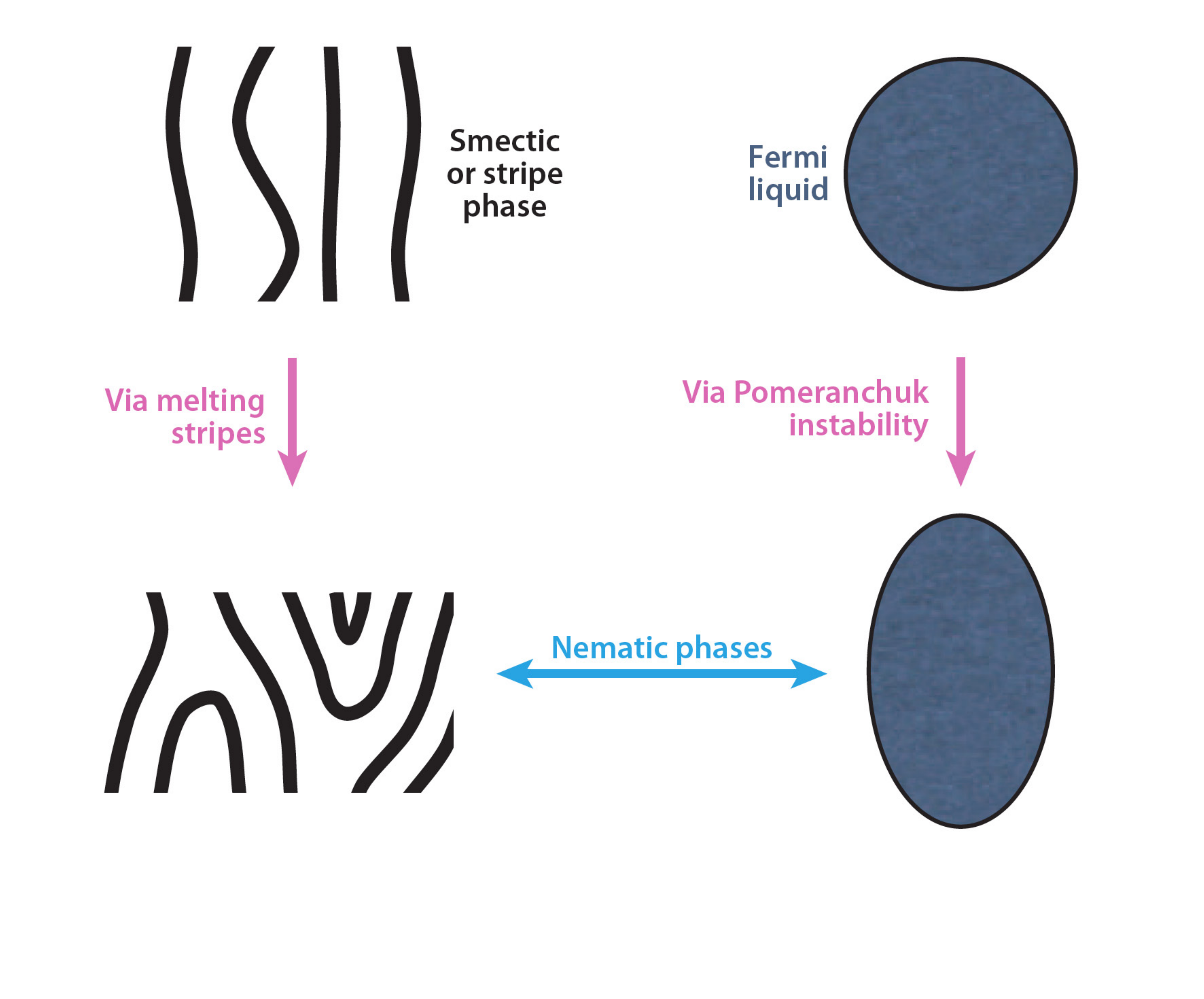}
\end{center}
\caption{Two different mechanisms for producing a nematic phase with point-particles:
The gentle melting of a stripe phase can
 restore long range translational 
 symmetry while preserving orientational order \cite{kivelson-1998}. 
 Alternatively, a nematic Fermi fluid can arise through the distortion of  the Fermi 
 surface of a metal via a Pomeranchuk instability \cite{oganesyan-2001,halboth-2000}. 
(After, in part, from Reference \cite{kivelson-1998}.)}
\label{fig:Mfig1}
\end{figure}

Nematic fluids can be visualized from either a strong coupling or a weak coupling perspective (see Figure \ref{fig:Mfig1}).  From the strong coupling perspective, one views a nematic as a partially melted solid. Specifically, a striped or smectic state can arise in various ways, and if it melts in such a way that the translational symmetry is restored but the orientation symmetry remains broken, the result is a nematic state.  Under certain circumstances, the thermal melting of a stripe state to form a nematic fluid is readily understood theoretically
\cite{fradkin-1999,sun-2008,berg-2009,radzihovsky-2008,zaanen-2004} and, indeed, the resulting description is similar to the theory of the nearly smectic nematic fluid that has been developed in the context of complex classical fluids \cite{chaikin-1995,nelson-1981}. Within this perspective, the nematic state arises from the proliferation of dislocations, the topological defects of the stripe state. This can take place either via a thermal phase transition (as in the standard classical case) or as a quantum phase transition. Whereas the thermal phase transition is well understood \cite{nelson-1981,chaikin-1995}, the theory of the quantum smectic-nematic phase transition by a dislocation proliferation mechanism is largely an open problem.
Two notable exceptions are the work of Zaanen et al. \cite{zaanen-2004} who studied this phase transition in an effectively insulating system,
and the work of Wexler \& Dorsey \cite{wexler-2001} who estimated the core energy of the dislocations of a stripe quantum Hall phase.
Nevertheless, the microscopic (or position-space) picture of the quantum nematic phase that results consists of a system of stripe segments, the analog of nematogens, whose typical size is the mean separation between dislocations. The system is in a nematic state if the nematogens exhibit long range orientational order on a macroscopic scale\cite{kivelson-1998,fradkin-1999}. However, since the underlying degrees of freedom are the electrons from which these nano-structures form, the electron nematic is typically an anisotropic metal. 
Similarly, nematic order can also arise from thermal or quantum melting a frustrated quantum antiferromagnet \cite{chandra-1990,capriotti-2004,read-1991}.

An alternative picture of the nematic state (and of the mechanisms that may give rise to it) can be gleaned from a
Fermi-liquid-like perspective. In this momentum space picture one begins with a metallic state consisting of a system of fermions with a Fermi surface (FS) and well defined quasiparticles (QP). In the absence of any sort of symmetry breaking the shape of the FS reflects the underlying symmetries of the system. 
There is a classic result due to Pomeranchuk \cite{pomeranchuk-1958} that shows that a Fermi liquid is thermodynamically stable provided none of the Landau parameters 
is negative and of large enough magnitude to overcome the stabilizing effects of Fermi pressure. 
When these conditions are violated a thermodynamic (Pomeranchuk) instability occurs and the system must undergo a quantum phase transition to a state in which the symmetries of the Fermi liquid state are lowered \cite{oganesyan-2001}. If the instability takes place in the spin-singlet channel,
 the result is that the FS spontaneously distorts. The shape of the FS in the stable broken symmetry phase depends on the angular momentum channel in which the instability takes place, on details of the quasiparticle dispersion relation, and on the many-body interactions (all of which are irrelevant to the physics of the Landau Fermi liquid). In its simplest version, the Pomeranchuk instability is a quantum phase transition in a 2D Fermi system between a Fermi liquid with a circular FS and a nematic Fermi fluid in which the FS has a quadrupolar distortion and hence looks like an ellipse. 

From the FS viewpoint, it is more straightforward to consider the nematic phase in terms of the anomalous expectation value of a fermion bilinear in the particle-hole channel, rather than the order parameters in Eq. \ref{N}.  For instance, in a rotationally invariant 2D system, we can define a Fermi surface distortion in a given angular momentum channel, $\ell \geq 1$, in terms of a complex order parameter $\mathcal{N}_\ell$:
\begin{equation}
\mathcal{N}_\ell=\sum_{\vec k} n(\vec k) \exp[i \ell \theta(\vec k)]
\end{equation}
where $n(\vec k)=\langle \psi^\dagger(\vec k) \psi(\vec k)\rangle$ is the expectation value of the occupation number for wavevector $\vec k$ 
 and $\theta(\vec k)$ is the 
polar angle defined by $\vec k$. 
One can define similar order parameters in 3D, labelled by the orbital angular momentum quantum numbers $(\ell,m)$ \cite{fregoso-2009}.
In the isotropic (Landau) phase $\mathcal{N}_\ell=0$ by symmetry. 
The {\em nematic} order parameter corresponds to $\ell=2$ ({\it i.e.\/} a quadrupolar distortion). 
Clearly, these particle-hole condensates 
are in some ways similar to the particle-particle condensates found in anisotropic superfluids, 
{\it e.g.} the p-wave superfluid $^3$He or the d-wave superconducting cuprates.
Indeed, Kee {\it et al} recently
 noted the existence of a unitary transformation
relating nematic order 
and d-wave superconductivity \cite{kee-2008}.
Similar
transformations relate d-wave superconductivity to either d-density wave order \cite{chakravarty-2001c},
or antiferromagnetism \cite{demler-2004}.
 
Other, still more exotic Pomeranchuk phases
are also possible. The $\ell=3$ state breaks parity and
 time-reversal symmetry, which makes it a continuum version of an orbital current loop phase \cite{sun-2008b,varma-2007}. An $\ell=6$ condensate corresponds to a hexatic phase.
If the instability occurs in a spin-triplet channel, phases that break both rotational invariance either in spin or in real space or both can occur. Examples of the latter are the nematic-spin-nematic phase \cite{kivelson-2003} 
and its generalizations \cite{wu-2007}.
In the spin triplet channel the QP's are subject to 
 an effective (and generally large) effective spin-orbit interaction resulting from the nature of the ordered state (and not from atomic physics).

 For a lattice system, the FS in the nematic state breaks the point-group symmetry.  The relevant fermion bilinear order parameters are essentially 
 an anisotropic component of the fermion kinetic energy. Several detailed microscopic studies of the mechanism of formation of a nematic phase have been carried out in the context of simple lattice models.  A
 nematic phase has been shown \cite{kivelson-2004} to be the ground state of the (three band) Emery model of the cuprates, in a carefully constructed strong coupling limit and close to half-filling.
A variety of lattice models have been studied, and have been shown to have nematic phases by means of perturbative approaches (and in some cases mean field theory) leading to a Pomeranchuk mechanism
\cite{halboth-2000,hankevych-2002,kee-2003,metzner-2003,khavkine-2004,neumayr-2003,yamase-2005,kee-2005,raghu-2009,puetter-2009,quintanilla-2006,lamas-2008a}. With the exception of systems near a van Hove singularity (vHs), the nematic quantum phase transition is generally found  at some finite value of a coupling constant (as in the Stoner theory of ferromagnetism), and it is typically 
found to be first order (although fluctuations can turn these transitions continuous \cite{jakubczyk-2009}). Although Hartree-Fock-type theories are not reliable in strong correlation regimes, they are useful in that they yield a simple and clear picture of the nematic phase. 

The most salient results from these theories of the  isotropic (Fermi liquid) quantum nematic phase transition include the following:
\begin{enumerate}
\item
 The anisotropic nature of the nematic ground state 
 leads to a transport anisotropy which is tuned by the magnitude of the order parameter; it becomes stronger deeper in the nematic phase. 
 \item
 At the nematic quantum critical point the collective modes
 are strongly (Landau) overdamped, with 
 dynamic quantum critical exponent $z=3$.
  In continuum systems, but not in lattice systems,
 the Goldstone modes continue to be similarly overdamped throughout the nematic phase, except for propagation along two symmetry determined directions where they are underdamped. 
 Where the collective modes are overdamped, they modify the leading low $T$ behavior of the specific heat. 
 In two dimensions, this contribution is found to scale as $T^{2/3}$. 
 (A recent reexamination of the quantum critical behavior at the nematic transition is given in References \cite{zacharias-2009} and \cite{metlitski-2010}.)
 In 3D the continuum model, the nematic phase is \cite{oganesyan-2001} a marginal Fermi liquid \cite{varma-1989}, and the specific heat scales at $T \ln T$ \cite{Millis1993}.
 \item
 Lattice effects generally gap the Goldstone modes, thereby suppressing the non-Fermi liquid behavior in the nematic ordered phase at very low energies. 
 Perturbative renormalization group methods show \cite{dellanna-2006} that
 at the quantum phase transition, lattice anisotropies are irrelevant 
 (although quantities such as quasiparticle rates are strongly anisotropic \cite{dellanna-2007}), thus restoring the non-Fermi liquid behavior of the continuum models sufficiently close to the transition.
 \item
 Perturbative approaches 
 predict a strong broadening of the QP excitations that in two dimensions acquire a decay rate $\Sigma^{''}(\omega)\sim \omega^{2/3}$, strong enough to wipe out the conventional QP pole of the Landau theory.  The same approaches, extended to the nematic phase in the continuum, result in a non-Fermi liquid phase with 
 a strongly anisotropic $\Sigma^{''}(\vec k,\omega)$ in momentum space: 
 It has the same singular frequency dependence as at the quantum critical point except when $\vec k$ is directed along a principle axis of the nematic state, where the QP's are still well defined.
 In contrast, in the nematic phase of lattice models at very low temperatures and frequencies, anisotropic Fermi liquid behavior is recovered below a low energy scale set by the magnitude of the crystal fields.
 Whereas all approaches confirm the breakdown of Fermi liquid theory over the entire Fermi surface at the quantum critical point, and for all but isolated points in the continuum nematic phase, the (non-perturbative) higher dimensional bosonization approach 
 predicts a still more drastic destruction of the quasiparticles \cite{lawler-2006}, leading to a form of local quantum criticality \cite{lawler-2007,chubukov-2005a}.
 It is worth to note that there are conflicting results on this subject. Chubukov et al. \cite{rech-2006} have argued that in itinerant quantum critical systems $\Sigma^{''} \sim \omega^{2/3}$ holds to all orders in perturbation theory. 
 (See, however, the results of Reference \cite{metlitski-2010}).
 \item
  Deep in the nematic phase, the electron nematic may subsequently undergo further symmetry breaking transition, such as
  to a smectic (stripe) phase. This quantum phase transition is \cite{sun-2008} a quantum mechanical analog of the 
  McMillan-deGennes nematic-smectic transition of classical liquid crystals \cite{degennes-1993,chaikin-1995}.
A signature of the vicinity of such a transition is the existence of (generally incommensurate) low energy collective modes associated with various density wave states: a fluctuating stripe state \cite{kivelson-2003}.
 \end{enumerate}

\section{The Quantum Hall Nematic Phase}
\label{sec:QHnematic}
\setcounter{footnote}{0}

Particularly striking evidence for the existence of an electron nematic phase comes from ultra-high mobility two-dimensional electron systems (2DES) in GaAs/GaAlAs heterostructures \cite{lilly-1999,du-1999}.  Such systems are of course best known for their display of the fractional quantized Hall (FQH)effect when a large magnetic field is applied perpendicular to the 2D plane. Unlike the FQH states
 which are primarily confined to the lowest Landau level, the nematic phases appear when the Fermi level lies near the middle of the third and higher Landau levels.  Additional interesting non-FQH collective phases also exist in these higher Landau levels, but they are not discussed here.

\subsection{Transport Anisotropy}

\begin{figure}[b]
\centering
\includegraphics[width=\textwidth]{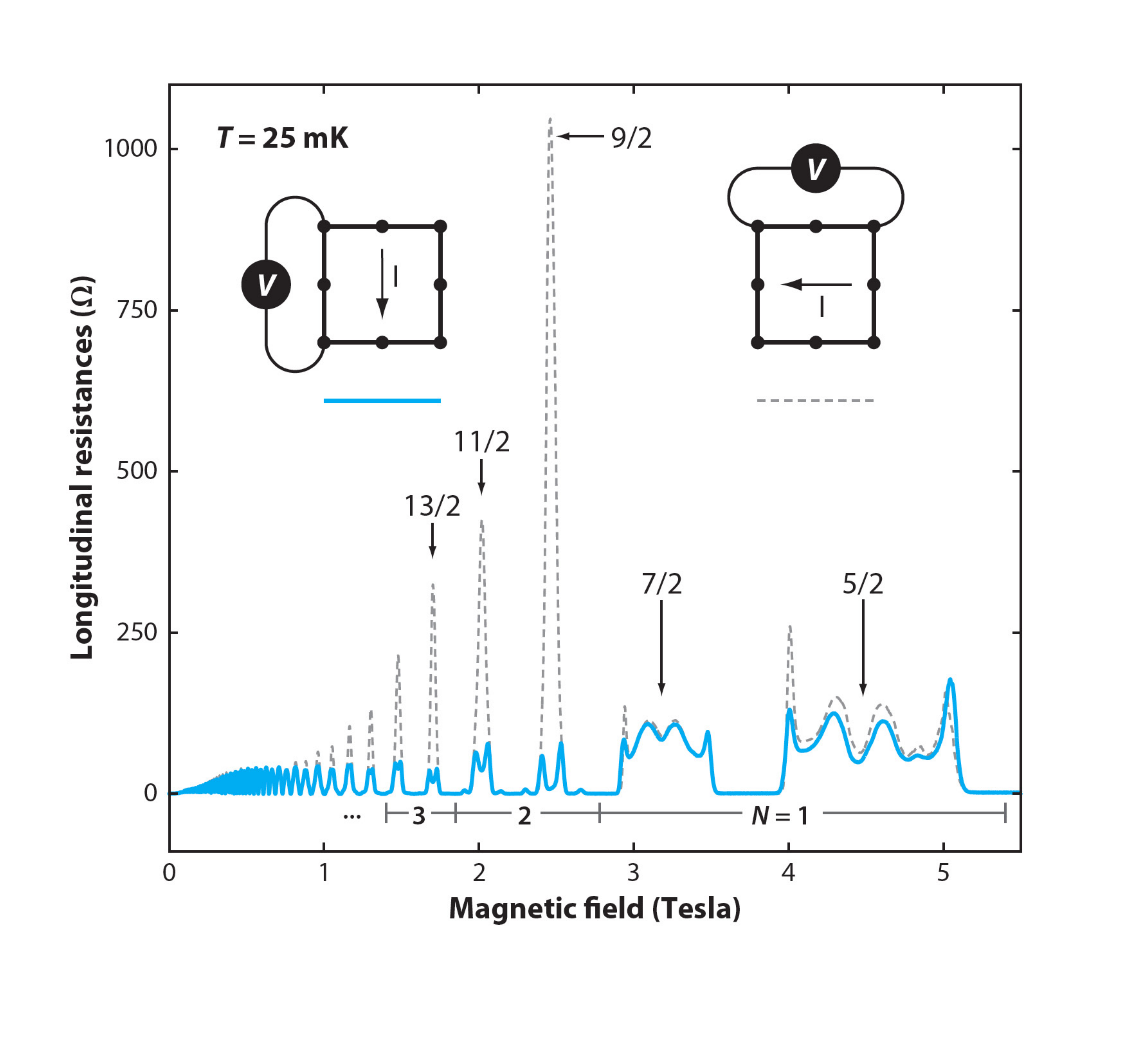}
\caption{\label{fig:Efig1} Longitudinal resistances in a high mobility 2D electron system at $T$ = 25 mK in a perpendicular magnetic field. Dashed gray line: current flow along \xdir.  Solid blue line: current flow along \ydir.  Strongly anisotropic transport is observed at Landau level filling factors $\nu = 9/2$, 11/2, 13/2, etc. No analogous anisotropy is present at very low magnetic fields or at fields above $B \approx 2.7$ T where the Fermi level falls into the $N=1$ Landau level.  [After Lilly {\it et al.} \cite{lilly-1999}.]}
\end{figure}

Figure \ref{fig:Efig1} shows typical low temperature magneto-resistance data from a high mobility ($\mu \sim 10^7$ cm$^2$/Vs) 2DES sample with a sheet density of approximately $n=2.7 \times 10^{11}$ cm$^{-2}$.  The sample itself is a square chip, 5 mm on a side, cleaved from a larger {\zdir}-oriented GaAs wafer upon which the epilayers needed to establish the 2DES were grown by molecular beam epitaxy.  Diffused Indium ohmic contacts are positioned at the corners and side midpoints of the square, as shown in the inset diagrams.  For the data shown in Figure \ref{fig:Efig1}  the magnetic field $B$ is perpendicular to the 2D plane. The Landau level filling fraction $\nu$ is determined by the 2D density $n$ and the spin-resolved level degeneracy $eB/h$.  For example, $\nu =nh/eB = 4$ at $B\approx2.7$ T. At higher magnetic fields the Fermi level lies in either the ground ($N=0$) or first excited ($N=1$) Landau level while at lower fields the Fermi level lies in the third ($N=2$) or higher Landau level.  The varying Landau level index $N$ is shown in the figure.

The two traces, dashed and solid, shown in Figure \ref{fig:Efig1} were taken during the same cool-down from room temperature.  The only difference between them is the mean direction of the current flow.  As the insets to the figure suggest, the current is driven between opposing mid-point contacts on the chip.  For the solid trace the mean current flow direction is along the \ydir\ crystallographic direction whereas for the dashed trace it is along \xdir. The associated voltage drops are measured between corner contacts as shown in the insets.  Remarkably, the resistances measured in these two orthogonal directions differ enormously at certain magnetic fields and yet are nearly equal at others.  In fact, the anisotropy shown in Figure \ref{fig:Efig1} is centered at half filling of several 
$N \geq 2$ Landau levels. At $\nu = 9/2$ in the $N=2$ Landau level the two resistances differ by a factor of approximately 100.  Substantial anisotropy is also observed at $\nu = 11/2$, 13/2, and 15/2, and a few higher half-odd integer filling factors before disappearing at higher $\nu$ in the low field semiclassical regime.  Note that no significant anisotropy is evident at $\nu = 5/2$ and 7/2 in the $N=1$ Landau level.
Interestingly, if the magnetic field is tilted away from perpendicular to the 2D plane, the otherwise isotropic FQHE states at $\nu = 5/2$ and 7/2 are destroyed and replaced by compressible anisotropic states \cite{pan-1999,lilly-1999b}.  Apparently, in a perpendicular field an electron nematic phase is only slightly higher in energy than these incompressible FQHE states and an in-plane field component can tip the balance. No analogous effect has been found anywhere in the $N=0$ Landau level.
[Nematic or hexatic versions of the FQH states may also exist, and were proposed sometime ago \cite{joynt-1996,balents-1996b}, but have not yet been observed.]
Though not shown in the figure, no evidence for any anisotropic states has been found within the $N=0$ lowest Landau level.  

\begin{figure}
\centering
\includegraphics[width=\textwidth]{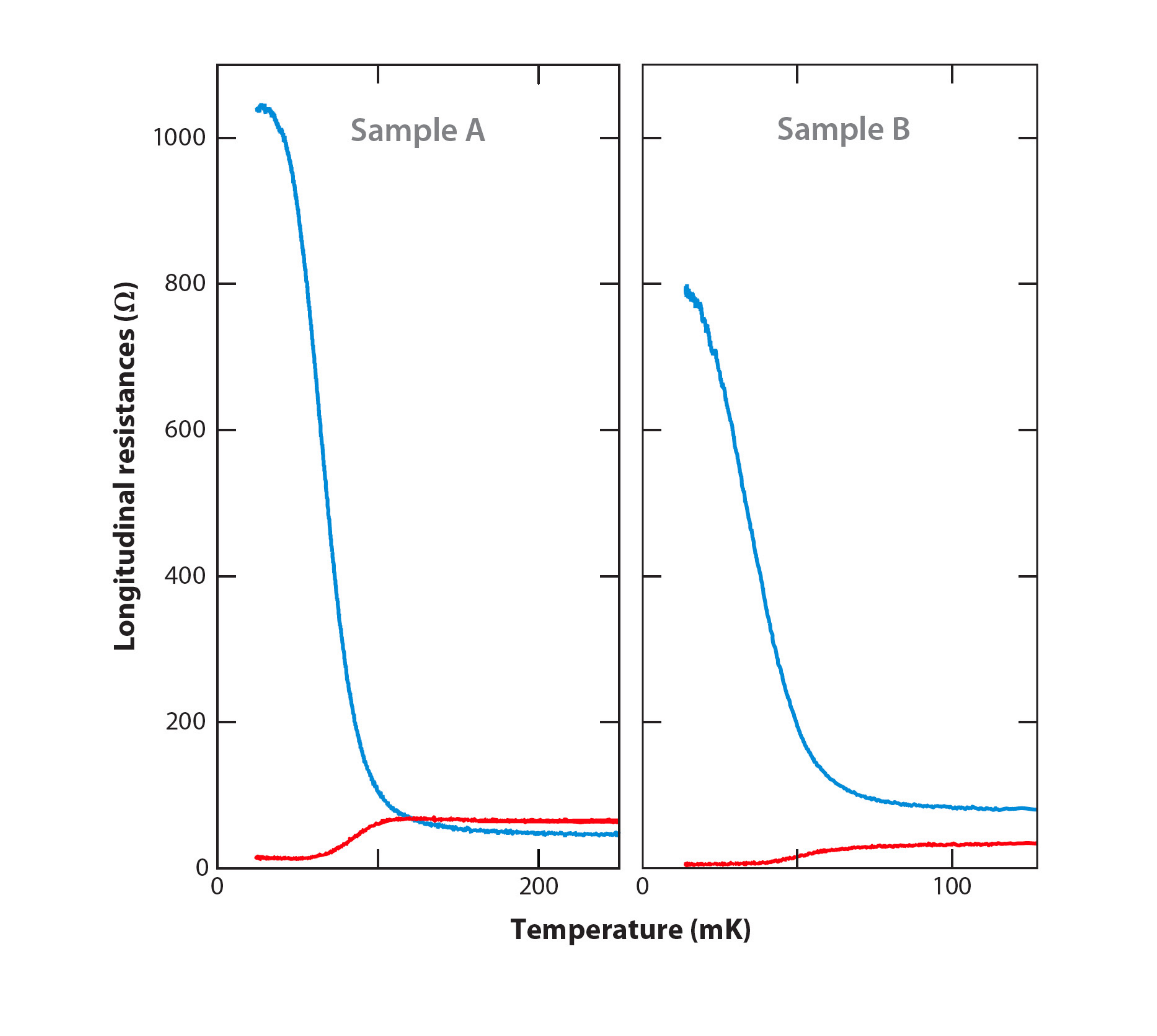}
\caption{\label{fig:Efig2} Temperature dependence of resistances along \xdir\ (blue line) and \ydir\ (red line) at $\nu = 9/2$ in two different samples.  The 2DES density of sample A is $n=2.67\times10^{11}$ cm$^{-2}$ whereas in sample B it is $n=1.5\times10^{11}$ cm$^{-2}$. [After Eisenstein at al. \cite{eisenstein-2001b}]}
\end{figure}

The anisotropy shown in Figure \ref{fig:Efig1} subsides rapidly with increasing temperature and is essentially absent above about $T = 150$ mK.  Figure \ref{fig:Efig2} shows the temperature dependence of the two resistances at $\nu = 9/2$ in two samples with different densities.  In both cases the anisotropy turns on rapidly at
 low temperatures.  Both samples also display a small resistive anisotropy (albeit of different signs) at high temperature.  This effect, which is largely independent of temperature and magnetic field, is commonly observed in GaAs/AlGaAs heterostructure samples and is generally attributed either to small misalignments of the contacts or to large scale density inhomogeneities. 
 It is readily distinguished from the sharply temperature and filling factor dependent anisotropy revealed in 
 Figure \ref{fig:Efig1} and Figure \ref{fig:Efig2}.

\subsection{Early Theory}

Hartree-Fock analyses of 2D electrons in high Landau levels by Koulakov at al. \cite{koulakov-1996,koulakov-1996b} and by Moessner \& Chalker 
\cite{moessner-1996},
found ground states with stripe order, unidirectional CDW states, due to the softening of the short range Coulomb interaction by the nodes of the electron wave functions in excited Landau levels.
For example, at $\nu = 9/2$ the system separates into equal width
stripes of $\nu = 4$ and $\nu = 5$.  Since the long-range part of the interaction is unaffected by the nodes, a maximum length scale $\lambda_{CDW}$ for the phase separation is imposed, and calculations suggest that $\lambda_{CDW} \sim R_c$, the classical cyclotron radius.  At $\nu = 9/2$ in typical samples, $\lambda_{CDW} \sim 100$ nm. In remarkable agreement with the subsequent experiments, these 
analyses \cite{koulakov-1996,koulakov-1996b} predicted that CDW states were favored 
in a range around 
half filling in the third and higher Landau level ({\it i.e.\/} at $\nu =9/2$, 11/2, etc.).  Quantum fluctuations in the ground and first excited Landau levels destroy CDW order and lead to uniform density liquid ground states.  Indeed, even in the $N=2$ and higher Landau levels quantum and thermal fluctuations can disrupt the long-range translational order of the CDWs at half-filling and render the system a nematic liquid crystal with short-ranged stripe order and long or quasi-long ranged
 orientational order \cite{fradkin-1999,fradkin-2000}.
 Far enough away from half filling, stripe-like order breaks down and crystalline bubble phases develop in the flanks of the Landau levels.  There is strong experimental evidence for these insulating bubble phases \cite{eisenstein-2001b}.

Not surprisingly, transport in 
 a striped system is expected to be highly anisotropic, with the quantum Hall edge state conductivity along the stripes greatly exceeding the hopping conductivity between stripes \cite{macdonald-2000,vonoppen-2000}.
Given that experimentally the hard transport direction lies along {\xdir}, we conclude that the stripes tend to lie along {\ydir}.

\subsection{Orientational Pinning}

The anisotropy of the resistance observed at $\nu = 9/2$, 11/2, etc. is consistently oriented relative to the crystalline axes: The ``hard'' direction is almost always along \xdir. 
[In some samples with very high electron density the hard direction switches to \ydir (see J. Zhu {\it et al.} \cite{Zhu-2002}).]
 Ultimately, there must exist a native symmetry breaking field within the semiconductor heterostructure that is responsible for this consistent orientation of the resistive anisotropy.  Remarkably, after a decade of research, this symmetry breaker has not been identified.  The crystal structure of bulk GaAs requires 4-fold symmetry (e.g. cos($4\phi$), with $\phi$ measured relative to, say, the \xdir\ direction) for the band structure of electrons confined to a {\zdir}-plane.  However, as Kroemer first noted \cite{Kroemer}, the 2DES lies close to one surface of a finite crystal and thus inversion symmetry is broken.  Although this opens the door to the needed 2-fold  (e.g. cos($2\phi$)) symmetry breaker, it does not identify it.

It has, however, turned out to be possible to estimate the strength of the ordering potential.  Early experiments \cite{pan-1999,lilly-1999b} showed that the application of a weak magnetic field component in the 2D plane could re-orient the anisotropy axes.  Comparison with detailed calculations \cite{jungwirth-1999,stanescu-2000} revealed that the orientational energy associated with the native symmetry breaking field is on the order of 1 mK per electron.  Interestingly, additional experiments \cite{cooper-2004} demonstrated that in addition to a native cos($2\phi$) symmetry breaking potential, there is also an even larger cos($4\phi$) potential, possibly arising from the piezoelectricity of GaAs \cite{fil-2000,rashba-1987}.

\subsection{Nematic to Isotropic Transition}

The early Hartree-Fock theories of CDW formation in high Landau levels estimated that the CDWs would form at temperatures of a few Kelvin \cite{koulakov-1996,koulakov-1996b}.  This is more than an order of magnitude larger than the typically 100 mK scale where resistive anisotropy begins to become pronounced at $\nu = 9/2$.  Although disorder in the 2D system might account for some of this suppression (in a manner similar to the reduction of FQH energy gaps), a more interesting possibility is that local stripe order may appear at quite high temperatures within small domains but thermal fluctuations prevent long range orientational order of those domains.  In this view, the resistive transition at approximately 100 mK is analogous to an isotropic to nematic phase transition in classical liquid crystals \cite{fradkin-1999,fradkin-2000,wexler-2001}.  

To examine this question, Cooper et al. \cite{cooper-2002} studied the temperature dependence of the resistive anisotropy at $\nu = 9/2$ in the presence of a large in-plane magnetic field.  Since an in-plane was already known to produce an orienting potential \cite{jungwirth-1999,stanescu-2000}, it 
is to be expected that if fluctuating stripe domains exist at temperatures above 100 mK the parallel field 
would order them and thereby induce resistive anisotropy at 
elevated temperatures.  

\begin{figure}
\centering
\includegraphics[width=\textwidth]{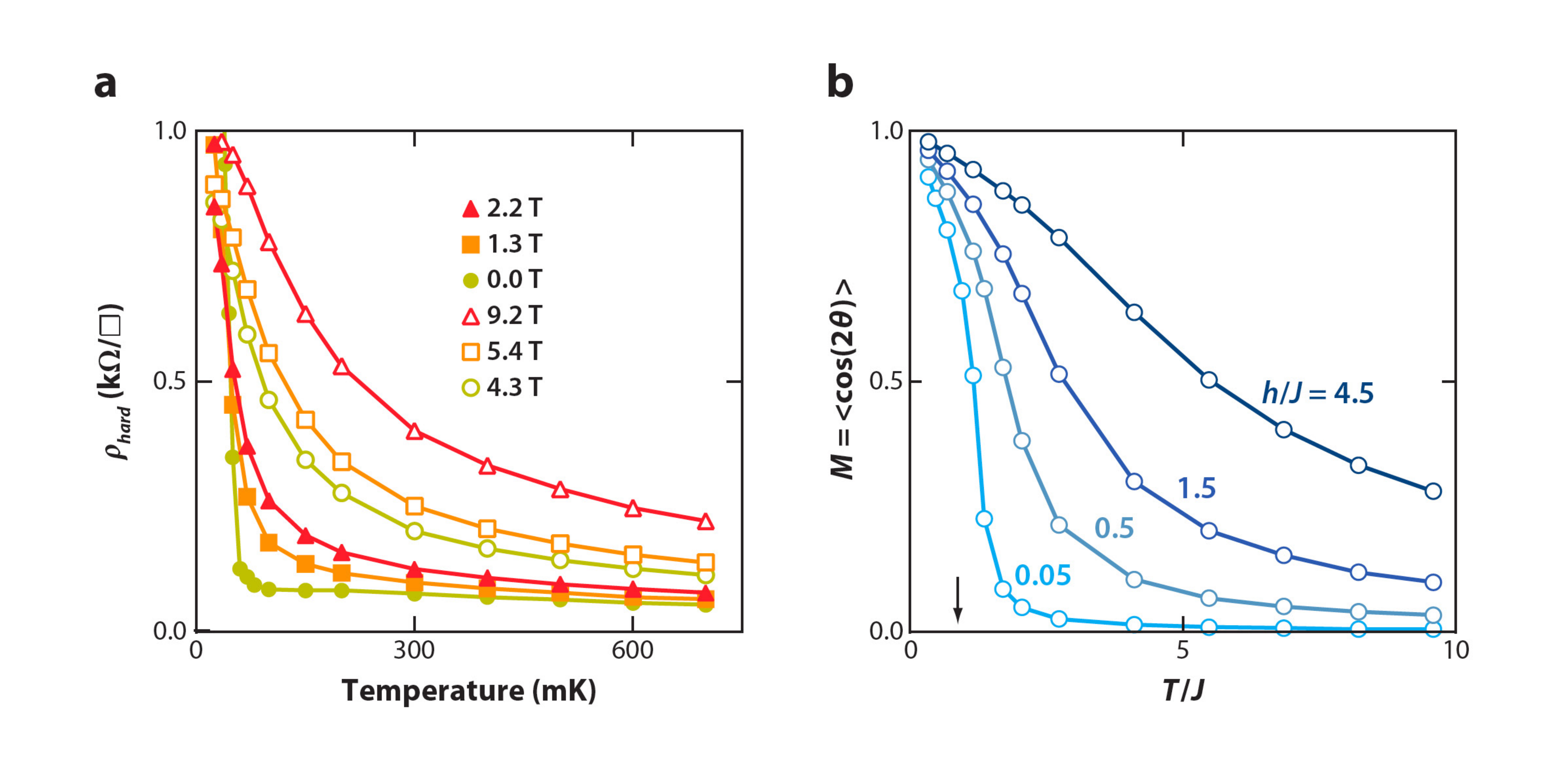}
\caption{\label{fig:Efig3} (a) Temperature dependence of $\rho_{\rm hard}$ at $\nu = 9/2$ for the various in-plane magnetic fields $B_{||}$ indicated in the legend.  (b) Temperature dependence of the order parameter of the 2D XY model for various symmetry breaking potentials $h/J$.  Arrow denotes Kosterlitz-Thouless transition temperature for $h=0$. [After Cooper et al. \cite{cooper-2002}]}
\end{figure}

In effect, this is precisely what Cooper et al. \cite{cooper-2002} observed. Figure \ref{fig:Efig3}(a) shows the temperature dependence of the resistivity.
The resistivity must be computed from the measured resistance taking into account 
geometric effects pointed out by S. Simon\cite{Simon-1999}.
$\rho_{hard}$ at $\nu =9/2$ in the hard {\xdir} direction for various in-plane magnetic fields applied along that same direction.
In-plane fields applied along the {\xdir} direction do not alter the orientation of the resistive anisotropy;
along {\ydir} they interchange the hard and easy transport directions, but 
produce the same expansion of the temperature scale of the anisotropy found for fields along {\xdir} \cite{cooper-2002}.
As observed previously, with no in-plane field applied (solid circles in Figure \ref{fig:Efig3}(a)) $\rho_{hard}$ is small and essentially temperature independent above about 100 mK while at lower temperatures it rises rapidly.  Adding an in-plane field gradually alters this temperature dependence, extending it to higher and higher temperatures.  With $B_{||}=9.2$ T applied along \xdir\ $\rho_{hard}$ remains quite substantial out to beyond 500 mK.  Meanwhile, since the resistivity in the easy direction shows relatively little dependence upon the in-plane field \cite{cooper-2002} we can conclude that the in-plane field has dramatically increased the temperature range over which resistive anisotropy exists at $\nu = 9/2$.
This result strongly supports the notion that local stripe domains exist at temperatures above 100 mK and that an in-plane field is merely orienting them. Importantly, in-plane field experiments performed at $\nu = 3/2$ in the lowest Landau level show no significant anisotropy at any similar temperature or in-plane field  (K.B. Cooper, J.P Eisenstein,
unpublished manuscript).  This discounts the idea that the in-plane field is itself $creating$ a significant anisotropy that does not exist in the absence of the field.

The in-plane field behavior discussed here is reminiscent of how a 
  ferromagnet responds to an external magnetic field.  In the absence of the external field the spins in the ferromagnet order spontaneously at the Curie temperature.  Above the Curie point the system is an ordinary paramagnet.  Applying a substantial magnetic field broadens the ferromagnetic transition, producing significant magnetization above the Curie point.  Following Fradkin et al. \cite{fradkin-2000}, Cooper et al. \cite{cooper-2002} performed Monte Carlo calculations on a classical 2D XY model with a varying $h \sim$ cos$(2\phi)$ symmetry breaking potential. As observed previously \cite{fradkin-2000}, such a model can simulate the high Landau level resistive anisotropy observed in purely perpendicular fields reasonably well.  Figure \ref{fig:Efig3}(b) demonstrates 
  that increasing the cos$(2\phi)$ term (which is assumed to be closely related to the in-plane magnetic field in the actual experiments) substantially extends the 
   range over which the 2D XY magnetization is large. The comparison of Figure \ref{fig:Efig3}(a) and Figure \ref{fig:Efig3}(b) is striking.

\section{Metamagnetism and nematicity in the bilayer ruthenate S\lowercase{r}$_3$R\lowercase{u}$_2$O$_7$}
\label{sec:SRO327}
\setcounter{footnote}{0}
\subsection{Basic properties and electronic structure}

The bilayer ruthenate metal Sr$_3$Ru$_2$O$_7$ is a member of the Ruddlesden-Popper series of layered ruthenates that
includes the itinerant ferromagnets SrRuO$_3$ \cite{Allen-1996} and Sr$_4$Ru$_3$O$_{10}$ \cite{Crawford-2002} and the unconventional superconductor Sr$_2$RuO$_4$ \cite{Mackenzie-2003}.  It is strongly two-dimensional, and can be thought of as a stack of weakly coupled bilayers, giving high conductivity in the $ab$ plane.  Although it does not order magnetically in ambient conditions, Sr$_3$Ru$_2$O$_7$ is an enhanced paramagnet with a Wilson ratio of order 10 \cite{Ikeda-2000}; charge and magnetic degrees of freedom are strongly coupled. 

\begin{figure}[h!]
\includegraphics[width=\textwidth]{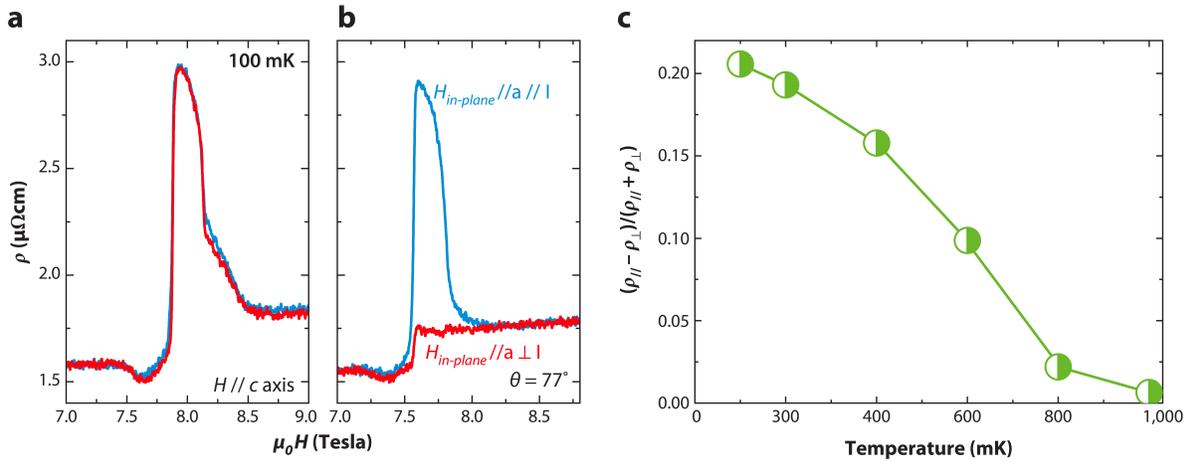}
\caption{The two diagonal components $\rho_{aa}$ and $\rho_{bb}$ of the in-plane magnetoresistivity tensor of a high purity single crystal of Sr$_3$Ru$_2$O$_7$.  (a)  For an applied field parallel to the crystalline $c$ axis (with an alignment accuracy of better than $2^\circ$), $\rho_{aa}$ (blue line) and $\rho_{bb}$ (red line) are almost identical.  (b) With the crystal tilted such that the field is $13^\circ$ from $c$, giving an in-plane component along $a$, a pronounced anisotropy is seen, with the easy direction for current flow being along $b$, perpendicular to the in-plane field component.   If the in-plane field component is aligned along $b$ instead, the easy direction switches to being for current flow along $a$.  c) The temperature dependence of the nematic order parameter defined by the resistive anisotropy for a tilt angle of $18^\circ$.  [After Borzi et al. \cite{borzi-2007}]}
\label{fig:Mackenzie1}
\end{figure}

The core experimental evidence for nematic fluid behavior in Sr$_3$Ru$_2$O$_7$ is shown in Figure \ref{fig:Mackenzie1}.  As a function of magnetic field applied parallel to the $c$ axis (i.e. perpendicular to the conducting layers), the material displays a standard metallic magnetoresistance up to approximately 7.5 T where it dips slightly before rising steeply at 7.8 T and then falling again at 8.1 T.  For the field applied in this high symmetry direction the measured in-plane resistivity remains four-fold symmetric throughout (Figure \ref{fig:Mackenzie1}a).  However, if the sample is tilted slightly to provide an in-plane field component, the resistivity becomes strongly anisotropic, with two-fold symmetry (Figure \ref{fig:Mackenzie1}b).  The resistively defined order parameter of Eq. 2.1 is shown in Figure \ref{fig:Mackenzie1}c, and is consistent with the behavior of a second-order phase transition in an external field conjugate to its order parameter.  

The natural interpretation of these data is that Sr$_3$Ru$_2$O$_7$ hosts a symmetry broken nematic phase between 7.8 T and 8.1 T, with domain formation masking the anisotropy when the field is perpendicular to the conducting planes.  Providing an in-plane field component aligns these domains and reveals the full effects of the symmetry breaking.

As described in Reference \cite{borzi-2007}, the phenomenology of nematic transport in Sr$_3$Ru$_2$O$_7$ is strikingly similar to that seen in the high purity 2DES discussed in Section \ref{sec:QHnematic} above, even though the underlying physics of the two systems appears to differ quite substantially.  Experimentally, they complement each other because each offers the opportunity to perform different classes of experiment.  Rather than repeat the discussion of transport given above, we will concentrate here on the new experiments that are accessible in Sr$_3$Ru$_2$O$_7$, notably bulk thermodynamics and surface spectroscopies.

The electronic structure of Sr$_3$Ru$_2$O$_7$ is complicated, with many bands crossing the Fermi level.  The experimental Fermi surface as determined by angle-resolved photoemission \cite{Tamai-2008},
is similar to the results of LDA calculations\cite{Singh-2001}.  As discussed by several authors \cite{Singh-2001, Tamai-2008, raghu-2009, Mercure-2009a}, much of the complexity results from an analysis that uses the relatively simple Fermi surface of the single layer material {\SROone}  
\cite{Bergemann-2003}, taking into account bilayer coupling and a $\sqrt{2}-\sqrt{2}$ reconstruction.  The symmetry lowering due to the reconstruction allows for a small pocket, centered on $\Gamma$, with Ru 4$d_{x^2-y^2}$ character, but this is not thought to have much influence on the metamagnetic or nematic physics.  The important sheets have strong Ru 4$d_{xz,yz}$ character ($\alpha_1$ and $\alpha_2$) or mixed 4$d_{xz,yz}$ and 4$d_{xy}$ character ($\beta$, $\gamma_1$, $\gamma_2$), with spin-orbit coupling also playing a role \cite{raghu-2009}.  Of particular significance to what follows are vHs centered on the tiny $\gamma_2$ pockets.  The $\gamma_2$ sheet could not be definitely determined to cross the Fermi level at the resolution of the photoemission study, but subsequent de Haas-van Alphen work gives evidence that it does \cite{Mercure-2009a}.  The de Haas-van Alphen measurements also confirm that the photoemission work in Sr$_3$Ru$_2$O$_7$ gives a genuine representation of the bulk properties, since there is good agreement between the two probes concerning both pocket areas and quasiparticle effective masses.

\subsection{Empirical link between metamagnetism and nematic phase formation}

\begin{figure}[b]
\includegraphics[width=\textwidth]{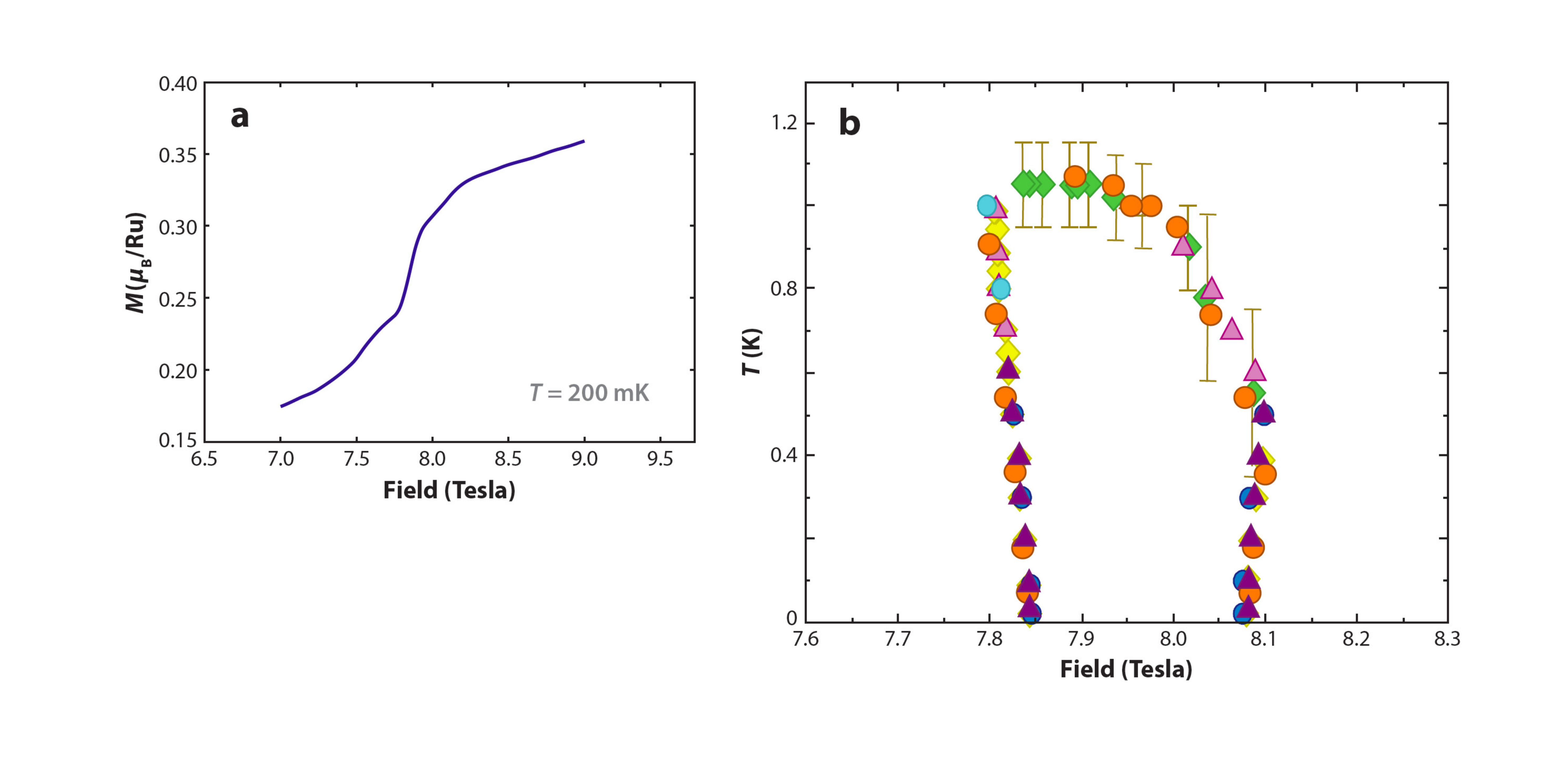}
\caption{a) Magnetization of Sr$_3$Ru$_2$O$_7$ as a function of magnetic field.  Three metamagnetic steps are seen.  The first, at 7.5 T, is a crossover, while the latter two are first-order phase transitions.  [After Rost et al. \cite{Rost-2009}].
b)  Low temperature phase defined from measurements of susceptibility, magnetostriction, thermal expansion and transport measurements. [After Grigera et al. \cite{Grigera-2004}].  Subsequent work established nematic transport properties within it \cite{borzi-2007} and the order of the phase transitions that define its boundaries \cite{Rost-2009}.
}
\label{fig:Mackenzie3}
\end{figure}

The coupling between spin and charge degrees of freedom in Sr$_3$Ru$_2$O$_7$ gives a natural link between nematicity and metamagnetism.  For magnetic fields applied parallel to the $c$ axis, samples with mean free paths of 300 {\AA}  have a single metamagnetic transition centered on approximately 7.7 T \cite{Perry-2001} and show no macroscopic nematicity.  If the purity is increased by an order of magnitude, this transition splits into three, as shown in Figure \ref{fig:Mackenzie3}a.  The rise in moment at 7.5 T is a crossover, but those at 7.8 and 8.1 T are first order phase transitions corresponding to the sharp features in resistivity shown in Figure \ref{fig:Mackenzie1} \cite{Perry-2004}.  Combining results from transport, thermal expansion, magnetostriction, a.c. susceptibility and magnetization allowed Grigera and co-workers to propose the existence of a fully bounded thermodynamic phase, as shown in Figure 6b \cite{Grigera-2004}.  A series of subsequent measurements have established the following key facts:

\begin{enumerate}
\item
Magnetocaloric and specific heat measurements show that all the boundaries in Figure \ref{fig:Mackenzie3}b represent transitions between equilibrium thermodynamic phases \cite{Rost-2009}.  
\item
 The transitions shown in Figure \ref{fig:Mackenzie3} are between metallic fluid phases.  This is confirmed both by transport  \cite{Perry-2004, Grigera-2004, borzi-2007} and, more recently, quantum oscillation measurements \cite{Mercure-2009b}.  The latter show that oscillations can be seen in each of the field ranges below, within and above the enclosed phase.
\item
In common with most itinerant metamagnets, there is a large magnetostriction, but within the resolution of elastic neutron scattering it is isotropic in the plane \cite{borzi-2007}.  This rules out the measured transport anisotropy being driven by a large structural change.
\item
 Unusually, the entropy within the phase is higher than that of the adjacent fluids.  Specifically, the entropy jumps up across the first-order phase transition at 7.8 T before falling again across that at 8.1 T.
\item
 The criterion for the observation of bulk nematicity is a mean free path of $\sim 1000$ {\AA} or more.  Fermi velocities in Sr$_3$Ru$_2$O$_7$ lie in the range 10 -- 50 km/s \cite{Tamai-2008, Borzi-2004}. Thus, this corresponds to a scattering rate of order 0.5 -- 3 K, spanning the observed $T_c$ of 1.2 K.  This disorder sensitivity is addressed in Reference \cite{Ho-2008} (though note the discussion in Section 2.2).
\end{enumerate}

\subsection{Angle dependent properties and the importance of spin-orbit coupling}

Although spin-orbit coupling in a ruthenate metal is expected to be weaker than that in, for example, a heavy-fermion system, it has been shown to play a measurable role in determining the electronic structure in Sr$_2$RuO$_4$ \cite{Haverkort-2008, Liu-2008}.  Its importance in Sr$_3$Ru$_2$O$_7$ is seen in the variation of the metamagnetic transition scale from $\sim 8$ T to $\sim 5$ T for applied fields $\parallel c$ and $\parallel ab$ respectively 
\cite{Perry-2001}.  In fact, rotation of the field between these two limits acts as a control parameter for the metamagnetism and nematicity, as shown first for slightly disordered samples \cite{Grigera-2003} and later for high purity samples displaying macroscopic nematicity \cite{Green-2005, borzi-2007}.  

The low temperature angle-dependent (the angle of the magnetic field with the $c$-axis) phase diagram of ultra-pure Sr$_3$Ru$_2$O$_7$
was presented in Reference \cite{Grigera-2003}.
As the angle increases, the region in the field-angle plane in which nematic transport is observed shrinks, remaining bounded by first-order phase transitions, until they merge into a single transition at approximately 40 degrees.  
There is also a second (and much less understood) phase that appears for fields applied near the $ab$ plane.
  It seems reasonable to attribute most of this phase richness to spin-orbit coupling, which will therefore have to be taken properly into account in any full theory of Sr$_3$Ru$_2$O$_7$, 
(see Reference \cite{raghu-2009}).

\subsection{Open questions}

In closing this section on {\SROtwo}, we describe a number of issues that have still to be resolved.  Some of these are specific to the particular situation in this material, whereas others are likely to apply to other nematic systems as well.

\subsubsection{The link to quantum criticality} 

Up to now, theoretical approaches to Sr$_3$Ru$_2$O$_7$ have been mean-field treatments, many stressing the significance of the vHs near the Fermi energy as a starting point for describing the metamagnetism and the nematicity \cite{Binz-2004, Grigera-2004, Green-2005, Berridge-2009, raghu-2009, lee-2009, Yamase-2007, Yamase-2007c, Puetter-2007}.  

\begin{figure}[h!]
\includegraphics[width=\textwidth]{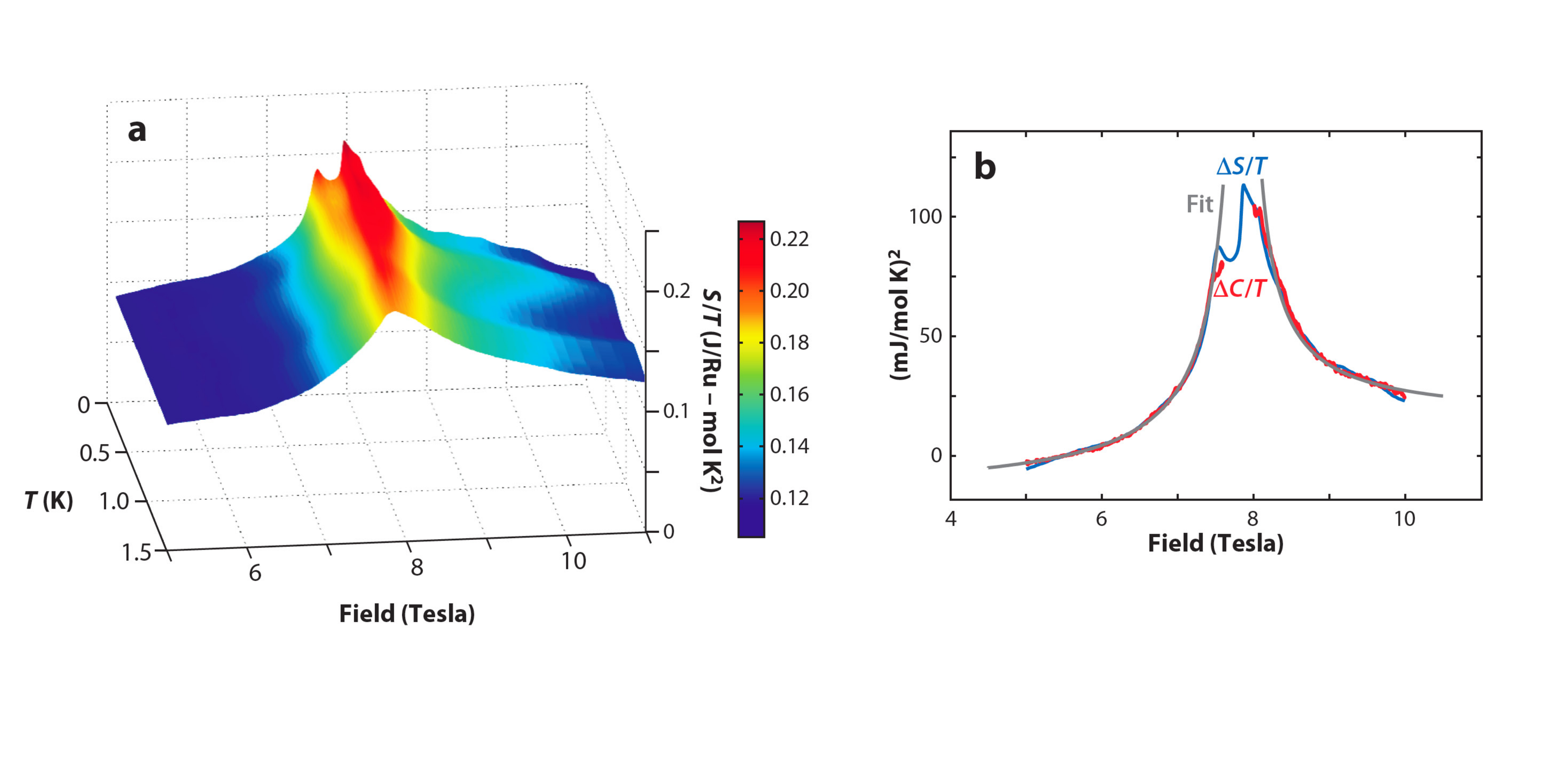}
\caption{a)  Entropy divided by temperature as a function of applied field and temperature for Sr$_3$Ru$_2$O$_7$.  b) Both $\Delta S/T$ (blue line) and the specific heat $\Delta C/T$ (red line) diverge as $\big|(H-H_c)/H_c\big|^{-1}$ outside the cut off region (grey lines). $\Delta C/T$ is fixed relative to $\Delta S/T$ at 5.5 T on the low-field side and 9.5 T on the high-field side.  [After Rost et al. \cite{Rost-2009}]
}
\label{fig:Mackenzie5}
\end{figure}

These vHs are certainly important, but several pieces of experimental evidence point to the need to go further.  First, experiments involving electron doping by substitution of La$^{3+}$ for Sr$^{2+}$ are inconsistent with the hypothesis of a simple rigid band shift of the chemical potential through fixed density of states peaks, even though the equivalent experiment in Sr$_2$RuO$_4$ is compatible with a rigid band shift \cite{Farrell-2008, Shen-2007}.   Second, the phase diagram of Sr$_3$Ru$_2$O$_7$ outside the metamagnetic region displays a number of transport and thermodynamic signatures of quantum criticality \cite{Perry-2001, Grigera-2001, Kitagawa-2005, Gegenwart-2006}.  This is summarized in the entropy landscape shown in Figure \ref{fig:Mackenzie5}a.  The entropy peak seen above the temperature of formation of the nematic phase is fully consistent with the approach to a quantum critical point.  At lower temperatures the peaking becomes sharper but is cut off with the onset of metamagnetism at 7.5 T.  The low temperature behavior can be put on a more quantitative footing by plotting the field dependence of $S/T$ and $C/T$ (Figure \ref{fig:Mackenzie5}b).  Below and above the metamagnetic phase both quantities diverge as $\big|(H - H_c)/H_c\big |^{-1}$.  Although vHs exist in Sr$_3$Ru$_2$O$_7$ and have, correctly, been emphasized in the 
(mean field) 
theories mentioned above, neither a simple one- or two-dimensional vHs is capable of capturing a divergence with this power law.   
 
A definitive interpretation of the thermodynamic data is still lacking.  The Hertz-Millis theory of the metamagnetic quantum critical point
\cite{millis-2002}
predicts a specific heat divergence as $\big|(H - H_c)/H_c \big|^{-1/3}$.  However, such theories calculate only the quantum critical contribution, whereas the experimental data are a combination of this and the 
mean-field 
contribution from the underlying density of states structure.  Work on relating critical fluctuations of a nematic to the behavior of Sr$_3$Ru$_2$O$_7$ has already been 
reported \cite{kao-2007}, and it seems clear that full theories going beyond the mean-field 
approximation will eventually be required.

\subsubsection{Entropy and specific heat in the nematic phase}

The entropic behavior of Sr$_3$Ru$_2$O$_7$ is also intriguing within the nematic phase itself.  First, the orientation of the first order transition lines in Figure \ref{fig:Mackenzie3} implies, through the application of the Clausius-Clapeyron law, that the low-temperature entropy is higher within the nematic phase than in the adjacent fluids.  Second, on cooling into the phase the specific heat shows the anomaly expected of a second order phase transition but the specific heat $C$ then contains a substantial quadratic correction to the temperature dependence expected in a Fermi liquid \cite{Rost-2009}.  Neither of these facts is currently understood.  

The entropy jump is surprising but not unprecedented.
For example, the Pomeranchuk effect in superfluid $^3$He is based on the existence of an enclosed phase with a higher entropy than that of its neighbors.
However, in {\SROtwo} the entropy jump should also be considered in the context of the broader entropy landscape.  As shown in Figure \ref{fig:Mackenzie5}b, the metamagnetic crossover at 7.5 T appears to cut off a divergence in $S/T$.  Although the entropy rises at the first-order transition into the nematic phase at 8.1 T, this is still against a background of a large saving of entropy compared to that which would have been developed had the divergence continued.  
It is tempting to think of domains as a source for the entropy rise in the nematic phase, and this possibility still needs to be investigated further.  However, the entropy jump on entering the phase at (7.8 T, 250 mK) (Figure 5b) is quite substantial: 5 mJ/molRuK or about one part in $10^3$ of $R\ln2$.  Another quantitative statement 
concerns the amount of entropy developed by $T_c$.  By the maximum $T_c$ of 1.2 K, S $\approx$ 250 mJ/molRuK, similar to the entropy developed by $T_c$ in many heavy-fermion superconductors \cite{Yang-2008c}, suggesting a possible entropic criterion for novel phase formation. 

\subsubsection{The scale of resistive anisotropy}

 A further outstanding challenge for theory is the size of the resistive anisotropy.  The `bare' anisotropy seen in Figure \ref{fig:Mackenzie1}b is nearly a factor of two, but it seems unlikely that all bands contribute equally to forming the nematic, thus some background 4-fold conductivity is to be expected.  The form of the two resistivity curves then suggests that more or less all of the extra scattering caused by entering the nematic phase is removed by driving the current along the easy direction.  No 
 weak coupling
 model involving point scattering is likely to be able to account for this using reasonable parameters.  A phase with $T_c \sim 1$K, i.e. a characteristic condensation energy of 0.1 meV, is 
 likely describable 
 at the weak-coupling end of the nematic spectrum as a Pomeranchuk distortion of the Fermi surface.  The low characteristic energy means that this distortion would be expected to be small,
 and, indeed, the absence of any detectable orthorhombic lattice distortion confirms this expectation.
 Even if it became Lifshitz-like due to the proximity of a vHs, it is hard to see how it could explain such a big anisotropy.  Moreover, quantum oscillation measurements indicate the existence of at least two closed sheets in the nematic phase, and an analysis of their amplitudes indicates an unusual length scale of order 500 nm for the onset of strong scattering \cite{Mercure-2009b}.  Taken together, the most natural picture to describe the resistive behavior seems to be strong scattering from domain walls \cite{Doh-2007} with a characteristic separation of approximately 500 nm.  Whatever the eventual explanation, the central challenge for theory is a plausible quantitative estimate of the scale of the scattering.

\section{Nematic phases in the cuprates and other interesting materials}
\label{sec:other}

Until now, we have focused our attention on the cleanest, clearest of experimental examples, and the simplest and most unambiguous of theoretical considerations.  In the final two sections of this review, we very briefly address more complicated, controversial, but in some ways even more exciting issues concerning the occurrence of nematic phases in some of the currently most interesting of materials. Along with this, we will briefly touch on some of the vexing issues associated with identifying nematic phases in more usual circumstances in strongly correlated systems, where  even relatively weak quenched disorder obscures most macroscopic manifestations of nematicity, and where the interplay between nematic order and other forms of order leads to bewildering complexities of the phase diagram.

\subsection{Nematic order in {\YBCO}}

Various striped phases, which is to say electron smectic phases, have been clearly identified in certain cuprate superconductors, especially those which are close relatives of {\LSCO}. 
In contrast, {\YBCO} is thought to be the cleanest of all the cuprate high temperature superconductors, and hence the place to  study the intrinsic clean limit physics; here the evidence of stripe order is more subtle, at best \cite{kivelson-2003}. However, there has recently accumulated increasingly compelling evidence that nematic order onsets in this material at a temperature which, at least in underdoped materials, is well above the superconducting $T_c$.  

{\YBCO} is orthorhombic, which is both good and bad news if we are interested in determining whether or not it is nematic:  The good news is that the orthorhombicity exerts a symmetry breaking field which tends to align nematic domains, making it possible to observe macroscopic anisotropy in various dynamical and thermodynamic properties of de-twinned single crystals.  The bad news is that some degree of macroscopic anisotropy is to be expected as a consequence of the orthorhombicity itself, unrelated to any interesting correlation effects.  The subtleties in the analysis of experiments, therefore, are associated with the need to distinguish anisotropies that result directly from the crystalline anisotropy, from those that are associated with the occurrence of a broken symmetry electronic state.
Two features of {\YBCO} make this possible.  First, there are no discernible structural changes that occur in the relevant range of temperatures, so any small and weakly $T$ dependent anisotropies seen at high temperatures can be  identified as the direct consequences of the crystal structure, while any large magnitude, strongly $T$ dependent enhancement of the anisotropy that occurs below a well defined crossover (transition) temperature, $T_N$, can be plausibly associated with the onset of electron nematic order.  (This is much the same logic as was applied in the analysis of the nematic ordering in the 2DES and in Sr$_3$Ru$_2$O$_7$).  Second, the crystals become increasingly orthorhombic with increasing doping, {\it i.e.} with increasing O concentration.  Thus, any contrary trend, where the low temperature electronic anisotropy increases with {\em decreasing} doping is plausibly associated with an  increasing tendency to nematic order in more underdoped crystals.  

Initial evidence for nematic order came from measurements of resistance anisotropy over a wide range of temperatures and doping concentrations
\cite{ando-2002}.  More recently, a heroic neutron scattering study \cite{hinkov-2007} of  the low-frequency and quasi-static magnetic structure factor of {\YBCO} with $x=6.45$ ($T_c=35$K) revealed order one anisotropy below a well-defined ordering temperature, $T_N \sim 150$K.  
 Most recently, a large anisotropy in the Nernst coefficient has been detected \cite{taillefer-nematic-2009} over a broad range of doping from slightly overdoped to very underdoped.  The Nernst anisotropy onsets below a doping dependent crossover temperature which follows the same general qualitative trends as the famous pseudo-gap temperature.  Taken together, these experiments offer strong evidence that electron nematic order underlies much of the pseudo-gap phase in {\YBCO}.

\subsection{Nematic order on the surface of {\BSCCO}}

{\BSCCO} apparently always has considerably more disordered than {\YBCO}. However, it cleaves at a mirror plane to reveal clean surfaces that have proven to be optimal for various forms of surface probes. Moreover, there are strong reasons to believe that the surface electronic structure is representative of that in the bulk.  In particular, STM studies of the surface electronic structure have yielded, among other things, vivid real-space pictures of various forms of local electronic structure. 

Initial studies identified \cite{howald-2003a,kivelson-2003} a nematic component to the surface electronic structure that was apparently related to a local tendency toward stripe order with a period near 4 lattice constants.  More recent studies of {\BSCCO} \cite{kohsaka-2007}  have greatly clarified the nature of this ordering:  There is a clear local tendency toward an electronically inhomogeneous state that strongly breaks the C$_4$ symmetry of the unit cell.
The local pattern is bond-centered, or equivalently is most apparent at the O sites.   It is stronger in more underdoped samples, although it is absent in samples that are sufficiently lightly doped that they are not superconducting.  It is glassy order, which is seen most clearly at energy scales comparable to the 
gap maximum and somewhat above, making some form of identification between this nematic order and the pseudo-gap natural.  Moreover, similar patterns have been seen \cite{hanaguri-2004} in STM studies of {\oxychloride} (which also has good surfaces for STM), making it clear that this is a generic property of cuprates. 

Because of the glassy character of the nematic order, it is not immediately clear how to best analyze it.  Various algorithms have been proposed \cite{robertson-2006,kivelson-2003,delmaestro-2006}
 for extracting the character of the underlying nematic and/or stripe correlations from the sort of glassy correlations that have been measured.  Very recently, an optimal approach has been developed for analyzing nematic correlations \cite{lawler-2009}, and this approach has been applied to data from underdoped {\BSCCO}.  The analysis leads to the conclusion that there is a reasonably long nematic correlation length, in excess of 100{\AA}.

\subsection{Nematic order in the Fe-pnictide superconductors}

Much as in the cuprates, the high temperature superconducting phase in the Fe-pnictides is generally derived by doping a  non-superconducting antiferromagnetic  ``parent'' material.  On doping, the antiferromagnetism is suppressed and eventually gives way to superconductivity, in some cases with an intermediate region of coexisting order.  Associated with the antiferromagnetism, there is a small structural distortion with nematic symmetry.  (In some of these materials, it is associated with a tetragonal to orthorhombic transition, and in others with an orthorhombic to monoclinic transition.)  That this distortion should be viewed as  a consequence of the development of electronic correlations, not as a structural artifact, follows from the now well established observation that it always follows similar trends as the antiferromagnetism, and vanishes at a critical doping concentration which, depending upon the material, either coincides with or lies close to the critical concentration at which antiferromagnetism disappears.  Indeed, more generally, the structural transition temperature always coincides with, or lies just slightly above the antiferromagnetic ordering temperature.  It thus appears that the structural distortion is a consequence of the development of an underlying electron nematic phase \cite{fang-2008,xu-2008,mazin}. 

\section{Some Future Directions}
\setcounter{footnote}{0}

There are many other correlated electronic systems in which nematic order could play a role.  In any system with orbital multiplicity at the Fermi energy, some form of orbital ordering is natural when interactions are sufficiently strong.  Since the Wannier functions associated with such bands typically transform non-trivially into each other under the point group operations, any orbitally ordered state necessarily breaks the point group symmetry and hence are nematic states.  Often, orbitally ordered states also break translational symmetry, but where they do not, orbital ordering and nematic ordering are, from a broken symmetry viewpoint, synonymous.  Indeed, the fact that the ruthenates and the Fe pnictides have such orbital multiplicities is probably one of the contributing factors in their nematic ordering.  Other transition metal oxides, such as the colossal magnetoresistive manganites, undoubtedly support nematic phases somewhere in their complex phase diagrams.  For instance, nematic order has also been see in the manganite material {\BiCaMnO} \cite{rubhausen-2000}. It has been suggested \cite{varma-2007} that the hidden order in {\URu2Si2} might be associated with a form of nematicity, where the very difficulties of detecting nematic order macroscopically may be responsible for its hidden nature.  The study of the many various ways such states can arise, and their differing effects on the properties of these materials is clearly a fertile area for future research.   

The role of orbital degeneracy in nematic order can produce a different type of nematic state in a quantum Hall device with a valley degeneracy, such as in AlAs \cite{shayegan-2008} or graphene \cite{castro-neto-2009}. Here, spontaneous valley ferromagnetism can produce a state with a nematic pattern of spatial symmetry breaking, but with vanishing dissipation and a quantized Hall conductance -- a quantized Hall nematic \cite{sondhi-2009}.  Theoretical arguments \cite{fradkin-1999,oganesyan-2001,spivak-2004b} have also been put forward suggesting the possibility of a quantum nematic or a quantum hexatic phase in the 2DES at zero magnetic field (and no disorder), near the border of the Wigner crystal phase.  
It has recently been  proposed that generalizations of quantum nematic phases may also occur in ultra-cold dipolar Fermi gases \cite{fregoso-2009,fregoso-2009b,quintanilla-2008,chan-2009} (atoms or molecules). 

A major area in which theoretical progress is desperately needed concerns the relation between transport properties and nematic order.  Clearly, much of the information about nematic phases comes from transport data.  Certain statements about the transport measurements follow from symmetry.
 However, beyond this, there has been relatively little  systematic study of transport in a nematic phase \cite{carlson-2006,vojta-2009,schmalian-2008}.  In particular, under circumstances in which the nematic domain size (which at low $T$ is determined by the character of the quenched disorder) is smaller than the normal state mean-free path, scattering of quasiparticles from the domain walls is likely to dominate the transport properties in ways that have not  begun to be theoretically explored.  

As already indicated, the nematic quantum critical point in disorder free systems is particularly interesting, as the critical fluctuations destroy the Fermi liquid character of the quasiparticle excitations over the entire Fermi surface.  Conceptually, this is one of the few places outside one-dimensional systems in which the breakdown of Fermi liquid theory can be seen in perturbation theory.  The character of the quantum critical state has already received some serious attention \cite{oganesyan-2001,lawler-2006,lawler-2007,chubukov-2004b,chubukov-2005a}, but more exploration of these points is warranted.  It would be even more exciting to study nematic quantum criticality in experiment -- for instance, there is some indication of the existence of an accessible nematic quantum critical point in transport experiments on {\LNSCO} \cite{Daou-2008b}.

There are even more exotic phases that are close relatives of the nematic phase, such as the nematic-spin-nematic phase, for which some suggestive \cite{fauque-2006}, but by no means conclusive evidence can be adduced.  From the perspective that ``whatever is not forbidden is compulsory," these phases have a right to exist, and we look forward to their identification, possibly in unexpected places, in the near future.

\section*{Disclosure Statement}

The authors are not aware of any affiliations, memberships, funding, or financial holdings that might be perceived as affecting the objectivity of this review.

\section*{Acknowledgments}
We thank E. Berg, H.-Y. Kee, E.-A. Kim, V. Oganesyan,  K. Sun,  J. Tranquada, and C. Wu  for great discussions. EF, MJL and SAK  thank the Kavli Institute for Theoretical Physics (KITP) of the University of California Santa Barbara for hospitality under the The Physics of Higher Temperature Superconductivity program.  This work was
supported in part by the National Science Foundation, under grants DMR
0758462 at the University of Illinois (EF), DMR 0531196 at Stanford University (SAK), PHY05-51164 at KITP (EF, MJL, SAK), and DMR-0070890 and DMR-0242946 at Caltech (JPE), and by the Office of Science, U.S.
Department of Energy, under contracts DE-FG02-07ER46453 through the Frederick
Seitz Materials Research Laboratory at the University of Illinois (EF),
DE-FG02-06ER46287 through the Geballe Laboratory of Advanced Materials at
Stanford University (SAK),  and DE-FG03-99ER45766 at Caltech (JPE), and by the UK Engineering and Physical Sciences Research Council under grant EP/F044704/1 (APM).

\providecommand{\newblock}{}


\end{document}